\documentstyle[preprint,prl,aps,epsfig]{revtex}
\tightenlines
\begin{document}
\draft

\title{Adaptation to the Edge of Chaos in the Self-adjusting Logistic Map}
\author{Paul Melby, J\"org Kaidel, Nicholas Weber, and Alfred H\"ubler}
\address{Center for Complex Systems Research, Department of Physics,
University of Illinois at Urbana-Champaign,
1110 West Green Street, Urbana, Illinois, 61801}

\date{\today}

\maketitle

\begin{abstract}

Self-adjusting, or adaptive, systems have gathered much recent 
interest.  We present a model for self-adjusting systems which treats 
the control parameters of the system as slowly varying, rather than 
constant.  The dynamics of these parameters is governed by a low-pass 
filtered feedback from the dynamical variables of the system.  We apply 
this model to the logistic map and examine the behavior of the control 
parameter.  We find that the parameter leaves the chaotic regime.  We observe
a high probability of finding the parameter at the boundary between
periodicity and chaos.
We therefore find that this system exhibits adaptation to the edge of chaos.

\end{abstract}

\narrowtext


Self-adjusting, or adapting, systems are ubiquitous in nature.  
Living systems are
constantly changing their own properties in response to their environment.
For example, the Asian firefly
will adjust its flashing frequency to match the frequency of outside
stimuli, such as the flashing of another firefly \cite{strogatz}.  
Because of the
importance of adaptation in natural systems, there have been many
studies which try to characterize evolutionary, adaptive behavior
\cite{crutchfield,bak}.  Many of
these studies show that adaptive systems will adapt to a new state at the
boundary of chaos and order, called the edge of chaos. 
N.H. Packard \cite{packard} showed
that this effect occurred
for a population of cellular automata rules evolving with a genetic
algorithm.  Pierre and H\"ubler \cite{pierre}  studied 
two competitive, adaptive agents which
used
both control and modeling to predict the behavior of the logistic map and
found that,
over time, the agents use a control which places the logistic map at the
edge of chaos.  The edge of chaos also occupies a prominent position because it
has been found to be not only an optimal setting for control of a 
system \cite{ritz}, but also an optimal 
setting under which a physical system can
support primitive functions for computation \cite{langton}.

We suggest a new model for self-adjusting systems, as proposed in Ritz and
H\"ubler \cite{ritz}.  An adjustable system
is a system where the control parameters of the system are not constant
in time.  Control parameters are distinguished from dynamical variables
through a separation of time scales, i.e. the control parameters vary
much more slowly than do the dynamical 
variables \cite{haken}.  The dynamics of the control
parameters is simple, over-damped motion without an attractor. 
If the forcing function
for the parameter depends only on the system itself, the system is called
self-adjusting.  We examine using a forcing function which is a low-pass
filtered feedback from the dynamical variables of the system.  We apply
this type of feedback to the logistic map and 
show that it exhibits adaptation to the
edge of chaos.

The logistic map has a dynamical variable $x_{n}$ and parameter $a$ and is
a function of time, $n$:
\begin{equation} \label{eq:logistic}
x_{n+1} = ax_{n}(1-x_{n})
\qquad 0 \leq x_{n} \leq 1
\qquad 0 \leq a \leq 4
\end{equation}
The parameter $a$ determines the type of dynamics which occurs for the
dynamical variable, $x_{n}$ \cite{jackson}.  
If $0 < a < 3$, then the dynamics of $x_{n}$
has a stable fixed-point attractor.  For intermediate values, $3 < a <
3.569$ the dynamics of $x_{n}$ is periodic. 
For $3.569 < a < 4$, the dynamics of $x_{n}$ is mostly chaotic.  There are,
however, values of the parameter in this range which lead to periodic
behavior of $x_{n}$.  These values are called periodic windows.  $a=3.84$
is contained in the well known period 3 window.  
The edge of chaos refers to values of
$a$ which lead to periodic (chaotic) behavior in $x_{n}$ and with only a small
change would lead to chaotic (periodic) $x_{n}$ dynamics.  Thus, values of $a$
which are very near to 3.569, or are very near to the periodic windows, are
at the edge of chaos.

If the parameter, $a$, changes slowly with time, as in:
\begin{equation} \label{eq:parameter}
a_{n+1} = a_{n} + \epsilon  f_{n}
\qquad 0 \leq a_{n} \leq 4 \qquad n = 0,1,2,\ldots
\end{equation}
where $f_{n}$ is a forcing function, and $\epsilon$ is a small constant,
the logistic map becomes an adjustable system.
If the forcing, $f_n$, is a function, $g$, of only the 
dynamical variable, $x_n$, the
logistic map is self-adjusting.  Because of the requirement of a separation
of timescales, a low-pass filter is a logical choice.  By damping out the
high frequency terms, both the requirements of overdamped motion and
separation of timescales can be achieved.  In addition, low-pass filters are
common in natural and experimental situations.

The low-pass filtering can be achieved in numerical simulations by 
a Fourier analysis of the time series for $x_{n}$.  If $N$ time steps are used,
the Fourier sine and cosine coefficients are given by:
$\beta_{n0} = (1/N) \sum_{t=0}^{N-1} x(t+n-N+1)$, 
$\alpha_{nk} = (2/N) \sum_{t=0}^{N-1} x(t+n-N+1) \sin(2 \pi k t/N )$, and
$\beta_{nk} =  (2/N) \sum_{t=0}^{N-1} x(t+n-N+1) \cos(2 \pi k t/N)$
for
$k=1,2,..., (N-1)/2$ 
where k is the frequency.  If N is odd, an extra term is needed:
$\beta_{n (N+1)/2} = (1/N) \sum_{t=0}^{N-1} x(t+n-N+1)\cos(\pi (N+1)t/N)$.

A low-pass filter with DC cutoff and a very low frequency cutoff would
keep only terms $\alpha_{n1}$ and $\beta_{n1}$.  The back 
transformation would then become:

\begin{equation} \label{eq:backtransform}
\bar{x}_{n} = \alpha_{n1} \sin(\frac{2 \pi n}{N}) + \beta_{n1} \cos(\frac{2 \pi
n}{N})
\end{equation}

If the forcing is only applied once every $N$ steps, and is evaluated 
when $n$ is a multiple of $N$,  $f_{n}$ becomes simply:

\begin{equation} \label{eq:forcing}
f_{n} = \left\{ \begin{array}{ll} \epsilon \bar{x}_{N} = \epsilon \beta_{n1} &
   \textrm{if $n=iN$} \\
   0 & \textrm{if $n \neq iN$} 
\end{array} \right. \qquad \textrm { $i=1,2,3,\ldots$}
\end{equation}

Numerical simulations of the self-adjusting logistic map were performed.
$N \gg 1$ and $\epsilon \ll 1$ were used to ensure a good separation of
timescales. 
Fig \ref{fig1} shows the time dependence of 3 
different initial parameter values.
For $a_{0}=3.5$, there is no change in $a$ with time.  The limiting
dynamics of $x_n$ when $a=3.5$ is periodic.  However, the
dynamics of $a$ for both the initial values $a_{0}=3.8$ and
$a_{0} = 3.9$ shows a ragged time dependence until a value of $a$ is
reached that leads to a periodic limiting dynamics for $x_n$.
The system leaves the chaotic regime and settles on a periodic
dynamics.  The limiting value of $a$ leads to periodic behavior in $x_n$.
However, only a small change in this limiting value is necessary to create
chaotic dynamics in $x_n$.  Therefore we say that the system has adapted to
the edge of chaos.

To illustrate adaptation, 300 initial values of the parameter, 
$a_{0}$, were taken evenly over the interval [3.4:4].
Fig. \ref{fig2} shows a histogram for the distribution of parameter values
 for two times, $n=0$ and $n=60000$.
As can be seen, the initial distribution is flat over the
interval [3.4,4].  At $n=60000$, the probability is very small
for values of the parameter, $a$, whose limiting dynamics is chaotic.
The probability is very high, however, for those 
values of $a$ which have a limiting dynamics which is periodic.
In most cases, the system has
evolved to the periodic windows of the system, which are labeled in the
figure.  This high probability at values of $a$ which are at the edge of
chaos is an alternative description of adaptation to the edge of chaos, and is
satisfied by this system.
Initial parameter values which lead to periodic behavior have not
changed. 
This observation leads to an approximation for the behavior 
of the low-pass filtered dynamics, $\bar{x}_n$:
\begin{equation} \label{eq:lowpass}
\bar{x}_{n}(a) \approx \left\{ \begin{array}{ll} \delta_{n} &
   \textrm{if $a$ leads to chaotic  $x$ dynamics}\\
   0 & \textrm{if $a$ leads to periodic $x$ dynamics}
\end{array} \right.
\end{equation}
where $\delta_{n}$ is a nonzero number.  Equation \ref{eq:lowpass} can be
understood in terms of the recurrence time and power spectrum of the $x_n$
dynamics of the logistic map.  Periodic behavior has, by definition, a
finite recurrence time.  This leads to a power spectrum which has a lowest
frequency $\omega_{0}$, which is proportional to the inverse of the recurrence time.
Chaotic dynamics, however, has an infinite recurrence time 
and thus its power spectrum does not have a lowest 
frequency component \cite{schuster}.
Therefore, if the cut-off frequency of the low-pass filter is $\omega_{c}$, a
condition on the low-passed dynamics can be made:
\begin{equation} \label{eq:cutoff}
\bar{x}_{n}(a) = 0 \qquad \textrm{if $\omega_{c} < \omega_{0}(a)$}
\end{equation}
where $\omega_{0}(a)$ is the lowest frequency of 
the $x_n$ dynamics with parameter
value $a$.

To understand the ragged time dependence in Fig. \ref{fig1}, we look at the
autocorrelation function, $C$ of the feedback, $f_n$:
\begin{equation} \label{eq:auto}
C(j) \equiv \langle f_{n} f_{n+j} \rangle = \frac{1}{S} \sum_{n=0}^{n=S} f_{n} f_{n+j}
\end{equation}
where $S$ is the number of time steps taken while the parameter is still
changing.  A random data set will be $\delta$-correlated, as in:
\begin{equation} \label{eq:corr}
\frac{C(j)}{C(0)} =
\left\{ \begin{array}{ll} 1 & j = 0 \\
       0 & j \neq 0
\end{array} \right.
\end{equation} 
A comparison of the autocorrelation function of a random data set and the
feedback, $f_{n}$ is shown in Fig. \ref{fig3}.  As can be seen, the
feedback is very nearly $\delta$-correlated, which means that the ragged time
dependence observed in Fig. \ref{fig1} is a diffusive, random walk motion.

We have shown that, for the self-adjusting logistic map, initially chaotic
states adapt to periodic states which are at the edge of chaos.  
Our model uses a low-pass filtered feedback from the dynamical variables to
the parameter of the system.  This approach is different from previous
studies which have used cellular automata, genetic algorithms, or neural
networks to drive the adaptation.   Our model is much
simpler, using only feedback for adaptation.  A simple, feedback-based
model is more applicable to many physical systems which do not have gene
codes or memories.    We feel that adaptation to the edge of
chaos is a generic property of systems with a low-pass filtered feedback, 
independent of both the form of the low-pass filter and the specific system
under study.   The low-pass filtered feedback exploits basic properties of
periodicity and chaos, and so adaptation toward the edge of chaos
should be a common property of such systems. 

This research was supported by the Office of Naval Research Grant no.
N00014-96-1-0335.

\begin{figure}
\caption{
Time dependence of 3 initial parameter values.  $a=3.5$
corresponds to periodic motion, while $a=3.8$ and $a=3.9$ correspond to
chaotic motion.  The final value of $a_{0}=3.8$ is $3.74$ (period 5) and
the final value of $a_{0}=3.9$ is $3.96$ (period 4.) For this simulation,
$N=20$ and $\epsilon = 0.1$ were used.
}
\label{fig1}
\end{figure}

\begin{figure}
\caption{
Distributions of parameter values between 3.4 and
4.  The initial distribution at $n=0$ is flat, and the final distribution at
$n=60000$ clearly shows a high probability at the edge of chaos, in the
periodic windows.  The period of each window is labeled above the
corresponding peak.  $N=20$ and $\epsilon = 0.1$ were used in this
simulation.
}
\label{fig2}
\end{figure}

\begin{figure}
\caption{
(a) shows the autocorrelation function for a random data set, while
(b) shows the autocorrelation function for the feedback, $f_{n}$.  Both
functions are normalized to $C(0)$.  $N=20$ and $\epsilon = 0.1$ were used.
}
\label{fig3}
\end{figure}

\end{document}